\documentclass[10pt,twocolumn,preprintnumbers,superscriptaddress,nofootinbib,aps,prd]{revtex4-2}

\usepackage{graphicx}
\usepackage{amsmath}
\usepackage{srcltx}
\usepackage[T1]{fontenc}
\usepackage{tgchorus} 
\usepackage{calligra} 

\usepackage[utf8]{inputenc}
\usepackage[colorlinks]{hyperref}
\usepackage{url}
\usepackage{times}
\usepackage{amssymb}

\newcommand{\lsim}{\lesssim}

\newcommand{\eq}[1]{Eq.~(\ref{#1})}
\newcommand{\Ed}{E_{\gamma_d}}

\newcommand{\ord}[1]{\mathcal{O}{(#1)}}
\newcommand{\beq}{\begin{equation}}
\newcommand{\eeq}{\end{equation}}
\newcommand{\bea}{\begin{eqnarray}}
\newcommand{\eea}{\end{eqnarray}}
\newcommand{\eps}{\varepsilon}

\newcommand{\dph}{\gamma_d}
\newcommand{\md}{m_{\dph}}

\newcommand{\appropto}{\mathrel{\vcenter{
  \offinterlineskip\halign{\hfil$##$\cr
    \propto\cr\noalign{\kern2pt}\sim\cr\noalign{\kern-2pt}}}}}

\begin{document}

\pagestyle{plain}

\title{\boldmath  Thermal Emission of Dark Photons from Earth's Core}

\author{Hooman Davoudiasl}
\email{hooman@bnl.gov} 
\affiliation{High Energy Theory Group, Physics Department \\ Brookhaven National Laboratory,
Upton, NY 11973, USA}


\begin{abstract}
	
Dark photons in the sub-eV regime may be produced by the Earth's hot core, representing a much less extreme environment than stellar cores.  We consider this possibility and estimate constraints on the kinetic mixing parameter $\varepsilon$ that governs dark photon coupling to charged particles, using Earth core cooling arguments, as well as dark matter direct detection bounds from SENSEI and DAMIC-M experiments.  Our estimates suggest that the current results from these experiments constrain new dark photon parameter space.  We also find that the proposed Oscura experiment may reach two to three orders of magnitude below existing bounds on $\varepsilon$, for dark photon masses $\sim 10^{-4}$~eV, depending on the assumed parameters characterizing the Earth core.

\end{abstract}
\maketitle

{\it Introduction.---} \begin{calligra}{\large What makes up the Universe remains mostly unknown.}\end{calligra}  In particular, decades of experimental and theoretical activities have not yielded any specific clues regarding the identity and fundamental properties of its dominant material component, {\it i.e.} dark matter (DM).  In tandem, searches for physics beyond the Standard Model (SM) have not resulted in any firm signals, posing a challenge to a number of key ideas and their associated DM candidates.  At this juncture, absent any empirical evidence and compelling theoretical guidance, casting a wide net over the space of possibilities seems warranted.

One possibility that has attracted significant attention in recent years is the assumption that DM is not an extension of the SM, but instead resides in an entirely different sector, one endowed with its own hidden forces and a multiplicity of particles.  Such a scenario may be motivated by analogy with the SM: if the visible world has many fundamental ingredients it seems reasonable to assume the same for its invisible counterpart.  Nonetheless, it is generally necessary to allow that the visible and dark sectors interact at some, albeit suppressed, level in order to produce DM and end up with a consistent cosmological history.

A simple and motivated way to induce a connection between the SM and the dark sector \cite{Arkani-Hamed:2008hhe} is through a ``dark photon" $\dph$ of a hidden $U(1)_d$ gauge interaction that kinetically mixes with the ordinary photon $\gamma$ \cite{Holdom:1985ag}.  This mixing, governed by a small parameter $\eps\ll 1$, is allowed at the renormalizeable level, which makes it a {\it portal} to the dark sector.  It is generally assumed that $\dph$ has a mass $\md \neq 0$, either from the Stueckelberg mechanism \cite{Stueckelberg:1938hvi} or possibly generated by a dark sector Higgs field condensation.  

More explicitly, as required by electroweak symmetry, one can assume the sequence of interactions  
\beq
\frac{\eps}{2\cos\theta_W}\,F_{d\mu\nu}B^{\mu\nu} 
\underbrace{\longrightarrow}_{\rm EWSB} \frac{\eps}{2}\, F_{d\mu\nu} F^{\mu\nu}, 
\label{eq:epsFdB}
\eeq
where $\theta_W$ is the weak mixing angle, and the field strength tensor is defined by 
$
Z_{\mu\nu} \equiv \partial_\mu Z_\nu - \partial_\nu Z_\mu\,,
$ with $Z_{\mu\nu} = F_{d\mu\nu},B_{\mu\nu},F_{\mu\nu}$, corresponding to  
the $\dph$, SM hypercharge, and $\gamma$ field strength tensors, respectively.  Upon electroweak symmetry breaking (EWSB), one finds a mixing between the dark and visible photons, as implied by the right-hand expression in \eq{eq:epsFdB}.  One can then show that $\dph$ couples to the SM through  
\beq
\eps\, e\, A_{d\mu} \, J^\mu_{\rm em}\,,
\label{eq:AdJ} 
\eeq
where $e$ is the electromagnetic coupling constant, $A_{d\mu}$ is the gauge field associated with $\dph$, and $J^\mu_{\rm em}$ is the electromagnetic  current.  Therefore, $\dph$ interacts with all electrically charged particles in the manner that a photon does, but with a strength suppressed by $\eps$. 

Given the above universal interaction, $\dph$ can be probed in a variety of settings, using charged SM states.  These include both laboratory and astrophysical observations, over a wide range of $\md$ and $\eps$ (see, {\it e.g.}, Ref.~\cite{Caputo:2026pdw} for a recent survey).  Nonetheless, much parameter space remains open which, absent any strong conceptual arguments, can provide targets for experimental investigation.  In particular, if $\dph$ is a light mediator for secluded DM annihilation \cite{Pospelov:2007mp}, wide ranges of $\md$ and $\eps$ can in principle be relevant for establishing the observed relic DM abundance.  Alternatively, $\dph$ can be DM itself.  

In this {\it Letter}, we point out that the hot core of the Earth is a potential source of light dark photons, in the mass range $\md\lsim \ord{\rm 0.1~eV}$.  The emitted $\dph$ flux can travel to the Earth's surface without any significant depletion.  The core of the Earth is characterized by densities of $\ord{10~{\rm g\,cm}^{-3}}$ and temperatures $\lsim \ord{1~{\rm eV}}$ \cite{Davies2015CoreCooling}, which makes it a very different environment compared to astrophysical ones.  For example, the Solar core has a density of $\ord{100~{\rm g\,cm}^{-3}}$ and a temperature $\sim \ord{1~{\rm keV}}$ (see, {\it e.g.}, Ref.~\cite{Bahcall:2005va}).  Hence, the Earth can provide an interesting source, which sits between laboratory and astrophysical conditions \cite{Davoudiasl:2009fe}, though in principle stellar dynamics can yield stronger bounds.  Nonetheless, if the dark sector dynamics has significant environment dependence, for example caused by feeble long range interactions, constraints covering a  variety of sources can provide a more comprehensive probe of potential new phenomena.

Due to kinetic mixing with photons, the $\dph$ flux has a small probability $\propto \eps^2$ of converting into visible radiation along its trajectory (see, for example, Refs.~\cite{Mirizzi:2009iz,Jaeckel:2010ni}), which could be detected by a sufficiently sensitive photo-detector.  Since the $\dph$-$\gamma$ conversion can occur in vacuum, a potential search looking for this effect does not require any special medium or strong electromagnetic fields.  

In the following, we will focus on a different mechanism for detecting very light dark photons, one based on their {\it absorption} by electrons \cite{Dimopoulos:1985tm,Avignone:1986vm,An:2014twa,Bloch:2016sjj,Hochberg:2016sqx}.  This possibility has been considered in relation to direct detection of dark matter composed of $\dph$.  Also, $\dph$ can escape from the Earth core and contribute to its cooling, which can provide a constraint on the model.  A similar consideration was used in Ref.~\cite{Davoudiasl:2009fe}, but applied to light axion-like particles (one may also consider non-thermal processes to generate axion-like particles in the Earth \cite{ShekarWorkInProgress}).

Next, we will outline some of the background information and the formalism required for obtaining our estimates.

{\it Earth core.---}  We will adopt a simple model of the Earth core, in order to illustrate the main features of the phenomena and highlight approximate bounds that can be obtained from Earth core cooling arguments and DM direct detection  experiments.  Our conclusions suggest that further refinements of our exploratory approach could be warranted, but those would be outside the scope of this work.  

For simplicity, we take the Earth core to be made of iron (Fe), ignoring the small contribution of other elements.  We will take the Earth's inner core, composed of solid Fe to extend to a radius of $r_{\rm in}\approx 1220$~km and be characterized by a temperature of $T_{\rm in}\approx 6000$~K.  The outer core, made of molten iron, is taken to reach a radius of $r_{\rm out}\approx 3480$~km.  We adopt a temperature of $T_{\rm out}\approx 5000~\text{K}\approx 0.43$~eV, as a mean value between the lower ($\sim 6000$~K) and upper edges ($\sim 4000$~K) of the outer core \cite{Davies2015CoreCooling}.  We note that more recent investigations seem to support slightly larger temperatures for the inner-outer core boundary \cite{Wu2024IronMelting}, but the approximate values taken here provide reasonable benchmarks for our exploratory  estimates.  

In the iron-core system, we can assume that electrons are the only mobile charge sources, responsible for the potential emission of dark photons.  Iron is a transition metal with the electronic configuration Fe: [Ar] $3d^6$ $4s^2$.  We take the nearly free $4s$ electrons to constitute a degenerate gas with Fermi energy $E_F \approx 10.34$~eV \cite{Nautiyal1985FermiSurfaceFe}.  The associated Fermi momentum is given by 
\beq
p_F = \sqrt{2 m_e E_F}\approx 3.3~\text{keV}\,, 
\label{eq:pF}
\eeq
with the electron mass $m_e \approx 0.511$~MeV.  This treatment ignores some of the complications related to the atomic structure of the hot core, but given that we consider only thermal emission, and not atomic transitions, it should give a fair approximation.     

In what follows, we will assume that the dark photon mass is due to a Stueckelberg mechanism and there is no dynamical Higgs scalar present in the theory.  The regime of parameters that we will consider corresponds to $\md \ll \omega_p$, where $\omega_p$ is the plasma frequency of the thermal medium. The rate of dark photon emission from a hot plasma (like the Sun's interior) was estimated in Refs.~\cite{An:2013yfc,An:2013yua}.  This treatment is not for a degenerate Fermi gas.  However, we argue that one can use their results, for the Fe thermal system of the Earth's core, using empirically extracted values for $\omega_p$.  We will expand on this approach further below.    

{\it Formalism.---} According to the analysis of Refs.~\cite{An:2013yfc,An:2013yua}, the power per unit volume $V$ for resonant thermal emission of $\dph$ is dominated by its longitudinal mode and given by 
\beq
\frac{dP_L}{dV}\approx \frac{1}{4\pi} \frac{\eps^2 \, \md^2 \omega_p^3}{e^{\,\omega_p/T}-1}\,.
\label{eq:dPdV}
\eeq
While the above equation is derived in the classical limit, we take it as a fair order-of-magnitude estimate for the power output by the hot iron core.  This may be a reasonable expectation, given that the resonant emission of $\dph$ can be thought of as the gradual conversion of thermal photons into dark photons \cite{An:2013yfc}.  We also note that the longitudinal polarization function $\Pi_L$ that enters the derivation of \eq{eq:dPdV} has the same leading behavior in the classical limit for $T/m_e \ll 1$ and in the degenerate limit for $v_F\ll 1$ \cite{Braaten:1993jw}, where the Fermi velocity is given by $v_F^2=2 E_F/m_e$ (note that the definition of $\Pi_L$ differs between Refs.~\cite{An:2013yfc} and \cite{Braaten:1993jw}).   

To use the expression in \eq{eq:dPdV}, we need a value for $\omega_p$.  We will adopt the experimentally derived values, which have been quoted in the literature as $\omega_p = 4.9$~eV \cite{KUMAR2007185} and more recently $\omega_p = 3.5$~eV  \cite{Butler2021OpticalIron}.  Given the spread, we will provide results for two values of $\omega_p = 3.5, 5.0$~eV.  This is warranted, since the projections for experiments vary across this range of $\omega_p$ significantly, as we will discuss later.  Nonetheless, we find that either choice of parameters allows for exploration of the hitherto open dark photon parameter space.   

Let us note here that with the above choices of $\omega_p$ one avoids adopting naive estimates that would not capture the physics of the iron core.  To see this, we observe that the expected electron number density $n_e$ given the value of $p_F$ in \eq{eq:pF} would be 
\beq
n_e = \frac{p_F^3}{3 \pi^2}\approx (1.1~\text{keV})^3,
\label{eq:ne}
\eeq 
which would apparently lead to a plasma frequency
\beq
\omega_p^2 = \frac{e^2 n_e}{m_e} \approx (15~\text{eV})^2\,.
\label{eq:omegap}
\eeq
The above value is much larger than the experimentally deduced ones.  The origin of this discrepancy can be traced to the effect of bound electrons in Fe on the screening of charge seen by the free Fermi sea electrons (see, for example, Refs.~\cite{PhysRevB.15.1719,Butler2021OpticalIron}).

The form of \eq{eq:dPdV} does not account for effects of charge screening, due to a Debye mass for the photon in a plasma.  For a degenerate Fermi sea, this effect is governed by Thomas-Fermi screening, whose inverse length is given by (see, {\it e.g.}, Ref.~\cite{Raffelt:1996wa}) 
\beq
\lambda_{\rm TF}^{-2} = \frac{3\,\omega_p^2}{v_F^2}\,,
\label{eq:lamTF}
\eeq 
Using the above parameters, we find that $\lambda_{\rm TF}^{-1} \lsim 1.3$~keV, which is similar to the momentum transferred to electrons by the thermal photons in the core $q_e \approx \sqrt{2 E_\gamma  m_e}\approx 1.2$~keV, assuming $E_\gamma \sim 1.3$~eV, characteristic of a thermal population at temperatures  $T\sim 5000$-$6000$~K.  In particular, for resonant emission characterized by energies of $\ord{\omega_p}$ in \eq{eq:dPdV}, we have $q_e\sim 2$~keV.  Hence, at the level of our approximations, we may expect the above screening not have a significant effect on our estimates.  Also, since $T/E_F\ll 1$ for the Earth core system, thermal effects on the physical parameters of Fe can be ignored in our treatment.

\begin{figure}[t]\vskip0.25cm	
	\includegraphics[width=\columnwidth]{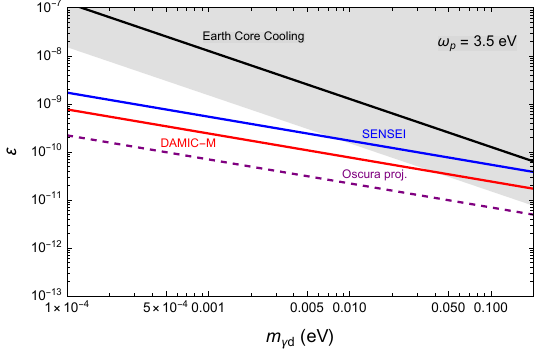}
	\caption{Constraints on and projections for $\eps$ as a function of dark photon mass $\md$ in eV, based on Earth core thermal emission rate estimates in this work.  The shaded gray region corresponds to an approximate functional fit of existing constraints (obtained with the help of ChatGPT) from various sources, compiled in Ref.~\cite{Caputo:2026pdw}, based on the data curated in Ref.~\cite{OHare2020AxionLimits}. From top to bottom, the lines  correspond to bounds from Earth cooling (solid black), recast SENSEI \cite{SENSEI:2023zdf} (solid blue) and DAMIC-M \cite{DAMIC-M:2025luv} experiments (solid red).  The last line is a projection for the proposed Oscura experiment \cite{Oscura:2022vmi} (dashed purple).  Here, we have set $\omega_p=3.5$~eV.}
	\label{fig:mainlow}
\end{figure}

\begin{figure}[t]\vskip0.25cm	
	\includegraphics[width=\columnwidth]{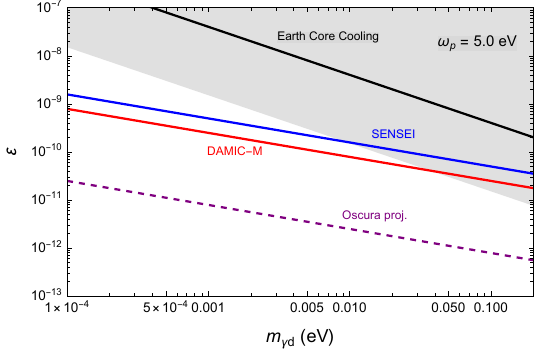}
	\caption{Same as Fig.\ref{fig:mainlow}, except that these results correspond to $\omega_p=5.0$~eV.}
	\label{fig:mainhigh}
\end{figure}

{\it Results.---} To get the total thermal $\dph$ output, we will sum over the inner and outer Earth core contributions, using the parameters adopted above.  We estimate that the total power emitted is
\beq
P_L^{\rm tot} \approx 63.1 \,(6.28) \times 10^{39}\, \eps^2 \left(\frac{\md}{\rm eV}\right)^2\, \text{erg s}^{-1}\,,
\label{eq:PLtot}
\eeq    
for $\omega_p = 3.5\, (5.0)$~eV.  Following the approach in Ref.~\cite{Davoudiasl:2009fe},  regarding {\it thermal geo-axion} production, we will use the heat flow estimates of the Earth core to constrain the production of dark photons.  The Earth core cooling rate can be approximated as \cite{Davoudiasl:2009fe} 
\beq
L_\oplus \sim 10^{19}~\text{erg s}^{-1}.
\label{eq:Core-cooling}
\eeq
We demand that $P_L^{\rm tot} < L_\oplus$, for a simple estimate of how geophysical considerations constrain the parameter space of the dark photon model. We have plotted the result from this bound in Figs.\ref{fig:mainlow} and \ref{fig:mainhigh} labeled by ``Earth Core Cooling" for the cases $\omega_p = 3.5, 5.0$~eV, respectively.  In these figures, the gray shaded area is ruled out by other data, as found in Ref.~\cite{Caputo:2026pdw}, which is based on the information in Ref.~\cite{OHare2020AxionLimits}.  The resulting constraints are weaker than the existing ones, as the figures show.  

Another venue for probing light dark photons, in the eV mass regime, is direct detection experiments.  In particular, one could recast the results of the experiments which look for eV scale dark matter bosons via absorption by electrons \cite{Dimopoulos:1985tm,Avignone:1986vm,An:2014twa,Bloch:2016sjj,Hochberg:2016sqx}.  The advent of skipper-CCD detectors \cite{Tiffenberg:2017aac} has opened a new front in the search for light dark matter and holds promise for reaching new regions of parameter space that have been hitherto inaccessible.  

The absorption of DM $\gamma_d$ (with negligible kinetic energy) is governed by the photo-electric cross section $\sigma_{\rm pe}(E_\gamma=\md)$ for photons, evaluated at a photon energy $E_\gamma=\md$.  The absorption rate per atom $R_{\rm abs}$ of DM $\gamma_d$ is governed by \cite{Bloch:2016sjj}  
\beq
R_{\rm abs}\approx n_{\gamma_d}\,v_{\gamma_d}\,\sigma_{\gamma_d} 
\approx \eps^2\,\sigma_{\rm pe}(E_\gamma=\md)\,,
\label{eq:Rabs}
\eeq
where $n_{\gamma_d} \approx \rho_{\gamma_d}/\md$ is the number density and $\rho_{\gamma_d} \approx 0.3~\text{GeV cm}^{-3}$ is the DM energy density in the Galactic neighborhood of the Earth; $v_{\gamma_d} \approx 238~\text{km s}^{-1}$ is the DM velocity and $\sigma_{\gamma_d}$ is the $\dph$ absorption cross section.  We have chosen these values so that a comparison with experimental data \cite{SENSEI:2024yyt} would be straightforward (we note that the density of DM near the Solar System has been reported to be somewhat larger, $0.4~\text{GeV cm}^{-3}$, in Ref.~\cite{Salucci:2010qr}).

We will now estimate the reach of current and proposed experiments which employ skipper-CCD technology, to probe the absorption of $\gamma_d$ emitted by the Earth core, based on their expectations for dark matter detection.   Given the core temperatures of $T\sim$ 5000-6000~K, the typical energy of thermal photons $E_\gamma \lsim$ 1.3~eV.  However, for resonant emission we have $\Ed = \omega_p$ and we work in the regime where $\md\ll \Ed$.    

We can interpret the quantity 
\beq
F_{\gamma_d}(\md) = n_{\gamma_d}(\md)\,v_{\gamma_d}
\label{eq:Fgammad}
\eeq
in \eq{eq:Rabs} as a flux of non-relativistic DM dark photons.  Given our parameters, we choose  reference DM masses $\md = 3.5, 5.0$~eV, which is set by the resonant emission energy $\Ed = \omega_p$.  To recast the DM absorption experimental results for the case at hand, we will need the thermal flux $F_\oplus({\omega_p})$ of $\gamma_d$ emitted by the core of the Earth, at its surface.  Using Eq.(\ref{eq:PLtot}), we have 
\beq
F_\oplus (\omega_p) \approx \frac{39.4\, (3.92) \times 10^{51}}{A_\oplus}\, \eps^2 \left(\frac{\md}{\rm eV}\right)^2\, \frac{\text{eV}}{\omega_p} \text{s}^{-1}\,,
\label{eq:EarthFux}
\eeq  
for $\omega_p = 3.5 \,(5.0)$~eV, where $A_\oplus = 4 \pi R_\oplus^2 \approx 5.10 \times 10^{18}$~cm$^2$ is the surface area of the Earth, corresponding to the mean Radius $R_\oplus\approx 6370$~km.  Here, $\omega_p$ is the typical $\gamma_d$ energy.  To apply the experimental results for DM absorption to our thermal $\gamma_d$ flux, we then require 
\beq
\eps^2 \,F_\oplus({\omega_p}) = \eps_{\rm exp}^2 \,F_{\gamma_d}({\omega_p})\,,
\label{eq:recast}
\eeq  
where $F_{\gamma_d}(\omega_p)\approx 2.04 \, (1.43) \times 10^{15}$~cm$^{-2}$~s$^{-1}$ is the DM flux at a mass equal to $\omega_p = 3.5 \,(5.0)$~eV, and $\eps_{\rm exp}$ is the experimental constraint or projection for $\eps$ at the corresponding DM mass $\md$.  

The preceding analysis assumed that the photo-electric cross section can be applied in going from a massless photon, to both a non-relativistic dark photon (DM) and a relativistic one (where we have focused on the regime $\Ed \gg \md$, for core-emitted $\gamma_d$).  This may seem unwarranted since the absorbed massless photon wavefunction includes a factor $e^{i\vec{k}.\vec{r}}$, where $\vec{k}$ and $\vec{r}$ are its spatial momentum and coordinate vector, respectively.  In the case of DM, since $v_{\gamma_d}\sim 10^{-3}$, this factor can be well-approximated by unity.  For the relativistic cases, the effect of the spatial momentum of the absorbed particle is small compared to the momentum transferred to the electron $\sim\ord{\sqrt{m_e\, \Ed}}$, as along as the $\Ed \ll m_e$ \cite{Pospelov:2008jk}, which is the case for our analysis.   

Let us first consider the recent SENSEI Collaboration experimental bounds \cite{SENSEI:2023zdf,SENSEI:2024yyt}.  Here, we need the constraint for a DM mass that matches our dark photon energy, {\it i.e.} $\md = \Ed=\omega_p = 3.5\, (5.0)$~eV, for which SENSEI results yield $\eps_{\rm exp} \lsim 3 \, (0.8) \times 10^{-13}$ \cite{SENSEI:2023zdf}.  We find that the recent DAMIC-M results \cite{DAMIC-M:2025luv} provide $\eps_{\rm exp} \lsim 6 \, (2) \times 10^{-14}$, for roughly the same DM masses. The corresponding bounds are presented in Figs.\ref{fig:mainlow} and \ref{fig:mainhigh} as solid blue (SENSEI) and red (DAMIC-M) lines.  We see that these experiments already probe new parameter space that has so far been open, based on our recast bounds for the case of thermal $\gamma_d$ emitted by the Earth core.

Next, we will estimate the reach of future measurements, based on the projections provided in Ref.~\cite{Oscura:2022vmi} for Oscura, as it yields the best reach.  For the same reference masses as before ({\it i.e.}, equivalent $\omega_p$), one gets $\eps_{\rm proj} \lsim 500\, (2) \times 10^{-17}$.  We have plotted the approximate projections in the $(\md, \eps)$ plane in Figs.\ref{fig:mainlow} and \ref{fig:mainhigh}, as dashed purple lines.  These projections suggest that significant new parameter space can be reached by an experiment that matches the capabilities proposed for Oscura, based on the reinterpretation of its DM direct detection data.  Note that the marked improvement of the reach for $\eps$ in going from $\omega_p = 3.5$~eV to $\omega_p = 5.0$~eV represents the projected enhanced sensitivity of the Oscura measurements at the latter DM mass value.  

Before closing, let us remark that the treatment presented here is an approximate one and a more detailed analysis is required for precise phenomenological results.  These could include a more elaborate  treatment of the Earth's core profile, its thermal and optical properties, and experimental considerations regarding DM direct detection.  Such refinements are left for future work, however our exploratory results can provide motivation for pursuing them. 

{\it Summary and conclusions.---} In this work, we considered the thermal emission of sub-eV dark photons from the Earth's core, which is characterized by less extreme conditions than stellar environments.  In this sense, our results represent thermal production in an intermediate regime, where possible departures from a laboratory setting are less significant.  We used Earth core cooling considerations as a means of constraining the relevant parameter space, which did not yield limits exceeding the prior ones.  Next, we used existing results from DM direct detection experiments and recast those to find new constraints on the dark photon parameter space.  Projections for a future experiment, with capabilities matching that of the Oscura proposal, seem to yield significant reach for currently open parameter space.  Our simplified treatment of the Earth's core and approximations used in calculating the emission and detection rates may need to be refined, using a more detailed analysis, in order to obtain more accurate constraints.  We, however, find our results sufficiently encouraging to motivate such elaborations, in future work.

\begin{acknowledgments}
We thank Rouven Essig for discussions and Maxim Pospelov for comments.  This work is supported by the US Department of Energy under Grant Contract DE-SC0012704.  ChatGPT was used in the production of Figs.\ref{fig:mainlow} and \ref{fig:mainhigh}, background material searches, and reference BibTeX formatting.  The author is responsible for the contents of this paper.  
\end{acknowledgments}

\begin{quote}
	{\small Digital data related to this work are included with the arXiv submission as ancillary files.}  
\end{quote}

\bibliography{geoDP.bib}

\begin{thebibliography}{33}%
\makeatletter
\providecommand \@ifxundefined [1]{%
 \@ifx{#1\undefined}
}%
\providecommand \@ifnum [1]{%
 \ifnum #1\expandafter \@firstoftwo
 \else \expandafter \@secondoftwo
 \fi
}%
\providecommand \@ifx [1]{%
 \ifx #1\expandafter \@firstoftwo
 \else \expandafter \@secondoftwo
 \fi
}%
\providecommand \natexlab [1]{#1}%
\providecommand \enquote  [1]{``#1''}%
\providecommand \bibnamefont  [1]{#1}%
\providecommand \bibfnamefont [1]{#1}%
\providecommand \citenamefont [1]{#1}%
\providecommand \href@noop [0]{\@secondoftwo}%
\providecommand \href [0]{\begingroup \@sanitize@url \@href}%
\providecommand \@href[1]{\@@startlink{#1}\@@href}%
\providecommand \@@href[1]{\endgroup#1\@@endlink}%
\providecommand \@sanitize@url [0]{\catcode `\\12\catcode `\$12\catcode
  `\&12\catcode `\#12\catcode `\^12\catcode `\_12\catcode `\%12\relax}%
\providecommand \@@startlink[1]{}%
\providecommand \@@endlink[0]{}%
\providecommand \url  [0]{\begingroup\@sanitize@url \@url }%
\providecommand \@url [1]{\endgroup\@href {#1}{\urlprefix }}%
\providecommand \urlprefix  [0]{URL }%
\providecommand \Eprint [0]{\href }%
\providecommand \doibase [0]{https://doi.org/}%
\providecommand \selectlanguage [0]{\@gobble}%
\providecommand \bibinfo  [0]{\@secondoftwo}%
\providecommand \bibfield  [0]{\@secondoftwo}%
\providecommand \translation [1]{[#1]}%
\providecommand \BibitemOpen [0]{}%
\providecommand \bibitemStop [0]{}%
\providecommand \bibitemNoStop [0]{.\EOS\space}%
\providecommand \EOS [0]{\spacefactor3000\relax}%
\providecommand \BibitemShut  [1]{\csname bibitem#1\endcsname}%
\let\auto@bib@innerbib\@empty
\bibitem [{\citenamefont {Arkani-Hamed}\ \emph {et~al.}(2009)\citenamefont
  {Arkani-Hamed}, \citenamefont {Finkbeiner}, \citenamefont {Slatyer},\ and\
  \citenamefont {Weiner}}]{Arkani-Hamed:2008hhe}%
  \BibitemOpen
  \bibfield  {author} {\bibinfo {author} {\bibfnamefont {N.}~\bibnamefont
  {Arkani-Hamed}}, \bibinfo {author} {\bibfnamefont {D.~P.}\ \bibnamefont
  {Finkbeiner}}, \bibinfo {author} {\bibfnamefont {T.~R.}\ \bibnamefont
  {Slatyer}},\ and\ \bibinfo {author} {\bibfnamefont {N.}~\bibnamefont
  {Weiner}},\ }\bibfield  {title} {\bibinfo {title} {{A Theory of Dark
  Matter}},\ }\href {https://doi.org/10.1103/PhysRevD.79.015014} {\bibfield
  {journal} {\bibinfo  {journal} {Phys. Rev. D}\ }\textbf {\bibinfo {volume}
  {79}},\ \bibinfo {pages} {015014} (\bibinfo {year} {2009})},\ \Eprint
  {https://arxiv.org/abs/0810.0713} {arXiv:0810.0713 [hep-ph]} \BibitemShut
  {NoStop}%
\bibitem [{\citenamefont {Holdom}(1986)}]{Holdom:1985ag}%
  \BibitemOpen
  \bibfield  {author} {\bibinfo {author} {\bibfnamefont {B.}~\bibnamefont
  {Holdom}},\ }\bibfield  {title} {\bibinfo {title} {{Two U(1)'s and Epsilon
  Charge Shifts}},\ }\href {https://doi.org/10.1016/0370-2693(86)91377-8}
  {\bibfield  {journal} {\bibinfo  {journal} {Phys. Lett. B}\ }\textbf
  {\bibinfo {volume} {166}},\ \bibinfo {pages} {196} (\bibinfo {year}
  {1986})}\BibitemShut {NoStop}%
\bibitem [{\citenamefont {Stueckelberg}(1938)}]{Stueckelberg:1938hvi}%
  \BibitemOpen
  \bibfield  {author} {\bibinfo {author} {\bibfnamefont {E.~C.~G.}\
  \bibnamefont {Stueckelberg}},\ }\bibfield  {title} {\bibinfo {title}
  {{Interaction energy in electrodynamics and in the field theory of nuclear
  forces}},\ }\href {https://doi.org/10.5169/seals-110852} {\bibfield
  {journal} {\bibinfo  {journal} {Helv. Phys. Acta}\ }\textbf {\bibinfo
  {volume} {11}},\ \bibinfo {pages} {225} (\bibinfo {year} {1938})}\BibitemShut
  {NoStop}%
\bibitem [{\citenamefont {Caputo}\ and\ \citenamefont
  {Essig}(2026)}]{Caputo:2026pdw}%
  \BibitemOpen
  \bibfield  {author} {\bibinfo {author} {\bibfnamefont {A.}~\bibnamefont
  {Caputo}}\ and\ \bibinfo {author} {\bibfnamefont {R.}~\bibnamefont {Essig}},\
  }\bibfield  {title} {\bibinfo {title} {{The Dark Photon: a 2026
  Perspective}}\ }(\bibinfo {year} {2026})\ \Eprint
  {https://arxiv.org/abs/2603.08430} {arXiv:2603.08430 [hep-ph]} \BibitemShut
  {NoStop}%
\bibitem [{\citenamefont {Pospelov}\ \emph
  {et~al.}(2008{\natexlab{a}})\citenamefont {Pospelov}, \citenamefont {Ritz},\
  and\ \citenamefont {Voloshin}}]{Pospelov:2007mp}%
  \BibitemOpen
  \bibfield  {author} {\bibinfo {author} {\bibfnamefont {M.}~\bibnamefont
  {Pospelov}}, \bibinfo {author} {\bibfnamefont {A.}~\bibnamefont {Ritz}},\
  and\ \bibinfo {author} {\bibfnamefont {M.~B.}\ \bibnamefont {Voloshin}},\
  }\bibfield  {title} {\bibinfo {title} {{Secluded WIMP Dark Matter}},\ }\href
  {https://doi.org/10.1016/j.physletb.2008.02.052} {\bibfield  {journal}
  {\bibinfo  {journal} {Phys. Lett. B}\ }\textbf {\bibinfo {volume} {662}},\
  \bibinfo {pages} {53} (\bibinfo {year} {2008}{\natexlab{a}})},\ \Eprint
  {https://arxiv.org/abs/0711.4866} {arXiv:0711.4866 [hep-ph]} \BibitemShut
  {NoStop}%
\bibitem [{\citenamefont {Davies}(2015)}]{Davies2015CoreCooling}%
  \BibitemOpen
  \bibfield  {author} {\bibinfo {author} {\bibfnamefont {C.~J.}\ \bibnamefont
  {Davies}},\ }\bibfield  {title} {\bibinfo {title} {Cooling history of earth's
  core with high thermal conductivity},\ }\href
  {https://doi.org/10.1016/j.pepi.2015.03.007} {\bibfield  {journal} {\bibinfo
  {journal} {Physics of the Earth and Planetary Interiors}\ }\textbf {\bibinfo
  {volume} {247}},\ \bibinfo {pages} {65} (\bibinfo {year} {2015})}\BibitemShut
  {NoStop}%
\bibitem [{\citenamefont {Bahcall}\ \emph {et~al.}(2006)\citenamefont
  {Bahcall}, \citenamefont {Serenelli},\ and\ \citenamefont
  {Basu}}]{Bahcall:2005va}%
  \BibitemOpen
  \bibfield  {author} {\bibinfo {author} {\bibfnamefont {J.~N.}\ \bibnamefont
  {Bahcall}}, \bibinfo {author} {\bibfnamefont {A.~M.}\ \bibnamefont
  {Serenelli}},\ and\ \bibinfo {author} {\bibfnamefont {S.}~\bibnamefont
  {Basu}},\ }\bibfield  {title} {\bibinfo {title} {{10,000 standard solar
  models: a Monte Carlo simulation}},\ }\href {https://doi.org/10.1086/504043}
  {\bibfield  {journal} {\bibinfo  {journal} {Astrophys. J. Suppl.}\ }\textbf
  {\bibinfo {volume} {165}},\ \bibinfo {pages} {400} (\bibinfo {year}
  {2006})},\ \Eprint {https://arxiv.org/abs/astro-ph/0511337}
  {arXiv:astro-ph/0511337} \BibitemShut {NoStop}%
\bibitem [{\citenamefont {Davoudiasl}\ and\ \citenamefont
  {Huber}(2009)}]{Davoudiasl:2009fe}%
  \BibitemOpen
  \bibfield  {author} {\bibinfo {author} {\bibfnamefont {H.}~\bibnamefont
  {Davoudiasl}}\ and\ \bibinfo {author} {\bibfnamefont {P.}~\bibnamefont
  {Huber}},\ }\bibfield  {title} {\bibinfo {title} {{Thermal production of
  axions in the Earth}},\ }\href {https://doi.org/10.1103/PhysRevD.79.095024}
  {\bibfield  {journal} {\bibinfo  {journal} {Phys. Rev. D}\ }\textbf {\bibinfo
  {volume} {79}},\ \bibinfo {pages} {095024} (\bibinfo {year} {2009})},\
  \Eprint {https://arxiv.org/abs/0903.0618} {arXiv:0903.0618 [hep-ph]}
  \BibitemShut {NoStop}%
\bibitem [{\citenamefont {Mirizzi}\ \emph {et~al.}(2009)\citenamefont
  {Mirizzi}, \citenamefont {Redondo},\ and\ \citenamefont
  {Sigl}}]{Mirizzi:2009iz}%
  \BibitemOpen
  \bibfield  {author} {\bibinfo {author} {\bibfnamefont {A.}~\bibnamefont
  {Mirizzi}}, \bibinfo {author} {\bibfnamefont {J.}~\bibnamefont {Redondo}},\
  and\ \bibinfo {author} {\bibfnamefont {G.}~\bibnamefont {Sigl}},\ }\bibfield
  {title} {\bibinfo {title} {{Microwave Background Constraints on Mixing of
  Photons with Hidden Photons}},\ }\href
  {https://doi.org/10.1088/1475-7516/2009/03/026} {\bibfield  {journal}
  {\bibinfo  {journal} {JCAP}\ }\textbf {\bibinfo {volume} {03}},\ \bibinfo
  {pages} {026}},\ \Eprint {https://arxiv.org/abs/0901.0014} {arXiv:0901.0014
  [hep-ph]} \BibitemShut {NoStop}%
\bibitem [{\citenamefont {Jaeckel}\ and\ \citenamefont
  {Ringwald}(2010)}]{Jaeckel:2010ni}%
  \BibitemOpen
  \bibfield  {author} {\bibinfo {author} {\bibfnamefont {J.}~\bibnamefont
  {Jaeckel}}\ and\ \bibinfo {author} {\bibfnamefont {A.}~\bibnamefont
  {Ringwald}},\ }\bibfield  {title} {\bibinfo {title} {{The Low-Energy Frontier
  of Particle Physics}},\ }\href
  {https://doi.org/10.1146/annurev.nucl.012809.104433} {\bibfield  {journal}
  {\bibinfo  {journal} {Ann. Rev. Nucl. Part. Sci.}\ }\textbf {\bibinfo
  {volume} {60}},\ \bibinfo {pages} {405} (\bibinfo {year} {2010})},\ \Eprint
  {https://arxiv.org/abs/1002.0329} {arXiv:1002.0329 [hep-ph]} \BibitemShut
  {NoStop}%
\bibitem [{\citenamefont {Dimopoulos}\ \emph {et~al.}(1986)\citenamefont
  {Dimopoulos}, \citenamefont {Starkman},\ and\ \citenamefont
  {Lynn}}]{Dimopoulos:1985tm}%
  \BibitemOpen
  \bibfield  {author} {\bibinfo {author} {\bibfnamefont {S.}~\bibnamefont
  {Dimopoulos}}, \bibinfo {author} {\bibfnamefont {G.~D.}\ \bibnamefont
  {Starkman}},\ and\ \bibinfo {author} {\bibfnamefont {B.~W.}\ \bibnamefont
  {Lynn}},\ }\bibfield  {title} {\bibinfo {title} {{Atomic Enhancements in the
  Detection of Weakly Interacting Particles}},\ }\href
  {https://doi.org/10.1016/0370-2693(86)91477-2} {\bibfield  {journal}
  {\bibinfo  {journal} {Phys. Lett. B}\ }\textbf {\bibinfo {volume} {168}},\
  \bibinfo {pages} {145} (\bibinfo {year} {1986})}\BibitemShut {NoStop}%
\bibitem [{\citenamefont {Avignone}\ \emph {et~al.}(1987)\citenamefont
  {Avignone}, \citenamefont {Brodzinski}, \citenamefont {Dimopoulos},
  \citenamefont {Starkman}, \citenamefont {Drukier}, \citenamefont {Spergel},
  \citenamefont {Gelmini},\ and\ \citenamefont {Lynn}}]{Avignone:1986vm}%
  \BibitemOpen
  \bibfield  {author} {\bibinfo {author} {\bibfnamefont {F.~T.}\ \bibnamefont
  {Avignone}, \bibfnamefont {III}}, \bibinfo {author} {\bibfnamefont {R.~L.}\
  \bibnamefont {Brodzinski}}, \bibinfo {author} {\bibfnamefont
  {S.}~\bibnamefont {Dimopoulos}}, \bibinfo {author} {\bibfnamefont {G.~D.}\
  \bibnamefont {Starkman}}, \bibinfo {author} {\bibfnamefont {A.~K.}\
  \bibnamefont {Drukier}}, \bibinfo {author} {\bibfnamefont {D.~N.}\
  \bibnamefont {Spergel}}, \bibinfo {author} {\bibfnamefont {G.}~\bibnamefont
  {Gelmini}},\ and\ \bibinfo {author} {\bibfnamefont {B.~W.}\ \bibnamefont
  {Lynn}},\ }\bibfield  {title} {\bibinfo {title} {{Laboratory Limits on Solar
  Axions From an Ultralow Background Germanium Spectrometer}},\ }\href
  {https://doi.org/10.1103/PhysRevD.35.2752} {\bibfield  {journal} {\bibinfo
  {journal} {Phys. Rev. D}\ }\textbf {\bibinfo {volume} {35}},\ \bibinfo
  {pages} {2752} (\bibinfo {year} {1987})}\BibitemShut {NoStop}%
\bibitem [{\citenamefont {An}\ \emph {et~al.}(2015)\citenamefont {An},
  \citenamefont {Pospelov}, \citenamefont {Pradler},\ and\ \citenamefont
  {Ritz}}]{An:2014twa}%
  \BibitemOpen
  \bibfield  {author} {\bibinfo {author} {\bibfnamefont {H.}~\bibnamefont
  {An}}, \bibinfo {author} {\bibfnamefont {M.}~\bibnamefont {Pospelov}},
  \bibinfo {author} {\bibfnamefont {J.}~\bibnamefont {Pradler}},\ and\ \bibinfo
  {author} {\bibfnamefont {A.}~\bibnamefont {Ritz}},\ }\bibfield  {title}
  {\bibinfo {title} {{Direct Detection Constraints on Dark Photon Dark
  Matter}},\ }\href {https://doi.org/10.1016/j.physletb.2015.06.018} {\bibfield
   {journal} {\bibinfo  {journal} {Phys. Lett. B}\ }\textbf {\bibinfo {volume}
  {747}},\ \bibinfo {pages} {331} (\bibinfo {year} {2015})},\ \Eprint
  {https://arxiv.org/abs/1412.8378} {arXiv:1412.8378 [hep-ph]} \BibitemShut
  {NoStop}%
\bibitem [{\citenamefont {Bloch}\ \emph {et~al.}(2017)\citenamefont {Bloch},
  \citenamefont {Essig}, \citenamefont {Tobioka}, \citenamefont {Volansky},\
  and\ \citenamefont {Yu}}]{Bloch:2016sjj}%
  \BibitemOpen
  \bibfield  {author} {\bibinfo {author} {\bibfnamefont {I.~M.}\ \bibnamefont
  {Bloch}}, \bibinfo {author} {\bibfnamefont {R.}~\bibnamefont {Essig}},
  \bibinfo {author} {\bibfnamefont {K.}~\bibnamefont {Tobioka}}, \bibinfo
  {author} {\bibfnamefont {T.}~\bibnamefont {Volansky}},\ and\ \bibinfo
  {author} {\bibfnamefont {T.-T.}\ \bibnamefont {Yu}},\ }\bibfield  {title}
  {\bibinfo {title} {{Searching for Dark Absorption with Direct Detection
  Experiments}},\ }\href {https://doi.org/10.1007/JHEP06(2017)087} {\bibfield
  {journal} {\bibinfo  {journal} {JHEP}\ }\textbf {\bibinfo {volume} {06}},\
  \bibinfo {pages} {087}},\ \Eprint {https://arxiv.org/abs/1608.02123}
  {arXiv:1608.02123 [hep-ph]} \BibitemShut {NoStop}%
\bibitem [{\citenamefont {Hochberg}\ \emph {et~al.}(2017)\citenamefont
  {Hochberg}, \citenamefont {Lin},\ and\ \citenamefont
  {Zurek}}]{Hochberg:2016sqx}%
  \BibitemOpen
  \bibfield  {author} {\bibinfo {author} {\bibfnamefont {Y.}~\bibnamefont
  {Hochberg}}, \bibinfo {author} {\bibfnamefont {T.}~\bibnamefont {Lin}},\ and\
  \bibinfo {author} {\bibfnamefont {K.~M.}\ \bibnamefont {Zurek}},\ }\bibfield
  {title} {\bibinfo {title} {{Absorption of light dark matter in
  semiconductors}},\ }\href {https://doi.org/10.1103/PhysRevD.95.023013}
  {\bibfield  {journal} {\bibinfo  {journal} {Phys. Rev. D}\ }\textbf {\bibinfo
  {volume} {95}},\ \bibinfo {pages} {023013} (\bibinfo {year} {2017})},\
  \Eprint {https://arxiv.org/abs/1608.01994} {arXiv:1608.01994 [hep-ph]}
  \BibitemShut {NoStop}%
\bibitem [{\citenamefont {Shekar}\ \emph {et~al.}(2026)\citenamefont {Shekar},
  \citenamefont {Dutta}, \citenamefont {Hu}, \citenamefont {Schneider},\ and\
  \citenamefont {Strigari}}]{ShekarWorkInProgress}%
  \BibitemOpen
  \bibfield  {author} {\bibinfo {author} {\bibfnamefont {A.~C.}\ \bibnamefont
  {Shekar}}, \bibinfo {author} {\bibfnamefont {B.}~\bibnamefont {Dutta}},
  \bibinfo {author} {\bibfnamefont {B.}~\bibnamefont {Hu}}, \bibinfo {author}
  {\bibfnamefont {A.}~\bibnamefont {Schneider}},\ and\ \bibinfo {author}
  {\bibfnamefont {L.}~\bibnamefont {Strigari}},\ }\href@noop {} {\bibinfo
  {title} {work in progress}} (\bibinfo {year} {2026})\BibitemShut {NoStop}%
\bibitem [{\citenamefont {Wu}\ \emph {et~al.}(2024)\citenamefont {Wu},
  \citenamefont {Wu}, \citenamefont {Wang}, \citenamefont {Ho}, \citenamefont
  {Wentzcovitch},\ and\ \citenamefont {Sun}}]{Wu2024IronMelting}%
  \BibitemOpen
  \bibfield  {author} {\bibinfo {author} {\bibfnamefont {F.}~\bibnamefont
  {Wu}}, \bibinfo {author} {\bibfnamefont {S.}~\bibnamefont {Wu}}, \bibinfo
  {author} {\bibfnamefont {C.-Z.}\ \bibnamefont {Wang}}, \bibinfo {author}
  {\bibfnamefont {K.-M.}\ \bibnamefont {Ho}}, \bibinfo {author} {\bibfnamefont
  {R.~M.}\ \bibnamefont {Wentzcovitch}},\ and\ \bibinfo {author} {\bibfnamefont
  {Y.}~\bibnamefont {Sun}},\ }\bibfield  {title} {\bibinfo {title} {Melting
  temperature of iron under the earth's inner core condition from deep machine
  learning},\ }\href {https://doi.org/10.1016/j.gsf.2024.101925} {\bibfield
  {journal} {\bibinfo  {journal} {Geoscience Frontiers}\ }\textbf {\bibinfo
  {volume} {15}},\ \bibinfo {pages} {101925} (\bibinfo {year}
  {2024})}\BibitemShut {NoStop}%
\bibitem [{\citenamefont {Nautiyal}\ and\ \citenamefont
  {Auluck}(1985)}]{Nautiyal1985FermiSurfaceFe}%
  \BibitemOpen
  \bibfield  {author} {\bibinfo {author} {\bibfnamefont {T.}~\bibnamefont
  {Nautiyal}}\ and\ \bibinfo {author} {\bibfnamefont {S.}~\bibnamefont
  {Auluck}},\ }\bibfield  {title} {\bibinfo {title} {Electronic structure of
  ferromagnetic iron: Fermi surface},\ }\href
  {https://doi.org/10.1103/PhysRevB.32.6424} {\bibfield  {journal} {\bibinfo
  {journal} {Physical Review B}\ }\textbf {\bibinfo {volume} {32}},\ \bibinfo
  {pages} {6424} (\bibinfo {year} {1985})}\BibitemShut {NoStop}%
\bibitem [{\citenamefont {An}\ \emph {et~al.}(2013{\natexlab{a}})\citenamefont
  {An}, \citenamefont {Pospelov},\ and\ \citenamefont {Pradler}}]{An:2013yfc}%
  \BibitemOpen
  \bibfield  {author} {\bibinfo {author} {\bibfnamefont {H.}~\bibnamefont
  {An}}, \bibinfo {author} {\bibfnamefont {M.}~\bibnamefont {Pospelov}},\ and\
  \bibinfo {author} {\bibfnamefont {J.}~\bibnamefont {Pradler}},\ }\bibfield
  {title} {\bibinfo {title} {{New stellar constraints on dark photons}},\
  }\href {https://doi.org/10.1016/j.physletb.2013.07.008} {\bibfield  {journal}
  {\bibinfo  {journal} {Phys. Lett. B}\ }\textbf {\bibinfo {volume} {725}},\
  \bibinfo {pages} {190} (\bibinfo {year} {2013}{\natexlab{a}})},\ \Eprint
  {https://arxiv.org/abs/1302.3884} {arXiv:1302.3884 [hep-ph]} \BibitemShut
  {NoStop}%
\bibitem [{\citenamefont {An}\ \emph {et~al.}(2013{\natexlab{b}})\citenamefont
  {An}, \citenamefont {Pospelov},\ and\ \citenamefont {Pradler}}]{An:2013yua}%
  \BibitemOpen
  \bibfield  {author} {\bibinfo {author} {\bibfnamefont {H.}~\bibnamefont
  {An}}, \bibinfo {author} {\bibfnamefont {M.}~\bibnamefont {Pospelov}},\ and\
  \bibinfo {author} {\bibfnamefont {J.}~\bibnamefont {Pradler}},\ }\bibfield
  {title} {\bibinfo {title} {{Dark Matter Detectors as Dark Photon
  Helioscopes}},\ }\href {https://doi.org/10.1103/PhysRevLett.111.041302}
  {\bibfield  {journal} {\bibinfo  {journal} {Phys. Rev. Lett.}\ }\textbf
  {\bibinfo {volume} {111}},\ \bibinfo {pages} {041302} (\bibinfo {year}
  {2013}{\natexlab{b}})},\ \Eprint {https://arxiv.org/abs/1304.3461}
  {arXiv:1304.3461 [hep-ph]} \BibitemShut {NoStop}%
\bibitem [{\citenamefont {Braaten}\ and\ \citenamefont
  {Segel}(1993)}]{Braaten:1993jw}%
  \BibitemOpen
  \bibfield  {author} {\bibinfo {author} {\bibfnamefont {E.}~\bibnamefont
  {Braaten}}\ and\ \bibinfo {author} {\bibfnamefont {D.}~\bibnamefont
  {Segel}},\ }\bibfield  {title} {\bibinfo {title} {{Neutrino energy loss from
  the plasma process at all temperatures and densities}},\ }\href
  {https://doi.org/10.1103/PhysRevD.48.1478} {\bibfield  {journal} {\bibinfo
  {journal} {Phys. Rev. D}\ }\textbf {\bibinfo {volume} {48}},\ \bibinfo
  {pages} {1478} (\bibinfo {year} {1993})},\ \Eprint
  {https://arxiv.org/abs/hep-ph/9302213} {arXiv:hep-ph/9302213} \BibitemShut
  {NoStop}%
\bibitem [{\citenamefont {Kumar}\ and\ \citenamefont
  {Auluck}(2007)}]{KUMAR2007185}%
  \BibitemOpen
  \bibfield  {author} {\bibinfo {author} {\bibfnamefont {M.}~\bibnamefont
  {Kumar}}\ and\ \bibinfo {author} {\bibfnamefont {S.}~\bibnamefont {Auluck}},\
  }\bibfield  {title} {\bibinfo {title} {Effect of pressure on the
  magneto-optical properties of bcc and bct iron},\ }\href
  {https://doi.org/https://doi.org/10.1016/j.physb.2006.08.012} {\bibfield
  {journal} {\bibinfo  {journal} {Physica B: Condensed Matter}\ }\textbf
  {\bibinfo {volume} {390}},\ \bibinfo {pages} {185} (\bibinfo {year}
  {2007})}\BibitemShut {NoStop}%
\bibitem [{\citenamefont {Butler}\ \emph {et~al.}(2021)\citenamefont {Butler},
  \citenamefont {Martinez}, \citenamefont {Kilci},\ and\ \citenamefont
  {Evans}}]{Butler2021OpticalIron}%
  \BibitemOpen
  \bibfield  {author} {\bibinfo {author} {\bibfnamefont {J.~K.}\ \bibnamefont
  {Butler}}, \bibinfo {author} {\bibfnamefont {M.}~\bibnamefont {Martinez}},
  \bibinfo {author} {\bibfnamefont {R.}~\bibnamefont {Kilci}},\ and\ \bibinfo
  {author} {\bibfnamefont {G.~A.}\ \bibnamefont {Evans}},\ }\bibfield  {title}
  {\bibinfo {title} {Optical properties of iron to 30 ev},\ }\href
  {https://doi.org/10.4236/msa.2021.1212042} {\bibfield  {journal} {\bibinfo
  {journal} {Materials Sciences and Applications}\ }\textbf {\bibinfo {volume}
  {12}},\ \bibinfo {pages} {622} (\bibinfo {year} {2021})}\BibitemShut
  {NoStop}%
\bibitem [{\citenamefont {Marton}\ and\ \citenamefont
  {Jordan}(1977)}]{PhysRevB.15.1719}%
  \BibitemOpen
  \bibfield  {author} {\bibinfo {author} {\bibfnamefont {J.~P.}\ \bibnamefont
  {Marton}}\ and\ \bibinfo {author} {\bibfnamefont {B.~D.}\ \bibnamefont
  {Jordan}},\ }\bibfield  {title} {\bibinfo {title} {Optical properties of
  aggregated metal systems: Interband transitions},\ }\href
  {https://doi.org/10.1103/PhysRevB.15.1719} {\bibfield  {journal} {\bibinfo
  {journal} {Phys. Rev. B}\ }\textbf {\bibinfo {volume} {15}},\ \bibinfo
  {pages} {1719} (\bibinfo {year} {1977})}\BibitemShut {NoStop}%
\bibitem [{\citenamefont {Raffelt}(1996)}]{Raffelt:1996wa}%
  \BibitemOpen
  \bibfield  {author} {\bibinfo {author} {\bibfnamefont {G.~G.}\ \bibnamefont
  {Raffelt}},\ }\href@noop {} {\emph {\bibinfo {title} {{Stars as laboratories
  for fundamental physics}: {The astrophysics of neutrinos, axions, and other
  weakly interacting particles}}}}\ (\bibinfo {year} {1996})\BibitemShut
  {NoStop}%
\bibitem [{\citenamefont {O'Hare}(2020)}]{OHare2020AxionLimits}%
  \BibitemOpen
  \bibfield  {author} {\bibinfo {author} {\bibfnamefont {C.}~\bibnamefont
  {O'Hare}},\ }\href@noop {} {\bibinfo {title} {Axionlimits: Repository of
  axion and dark photon constraints}},\ \bibinfo {howpublished}
  {\url{https://github.com/cajohare/AxionLimits}} (\bibinfo {year} {2020}),\
  \bibinfo {note} {accessed: May 2025}\BibitemShut {NoStop}%
\bibitem [{\citenamefont {Adari}\ \emph {et~al.}(2025)\citenamefont {Adari}
  \emph {et~al.}}]{SENSEI:2023zdf}%
  \BibitemOpen
  \bibfield  {author} {\bibinfo {author} {\bibfnamefont {P.}~\bibnamefont
  {Adari}} \emph {et~al.} (\bibinfo {collaboration} {SENSEI}),\ }\bibfield
  {title} {\bibinfo {title} {{First Direct-Detection Results on Sub-GeV Dark
  Matter Using the SENSEI Detector at SNOLAB}},\ }\href
  {https://doi.org/10.1103/PhysRevLett.134.011804} {\bibfield  {journal}
  {\bibinfo  {journal} {Phys. Rev. Lett.}\ }\textbf {\bibinfo {volume} {134}},\
  \bibinfo {pages} {011804} (\bibinfo {year} {2025})},\ \Eprint
  {https://arxiv.org/abs/2312.13342} {arXiv:2312.13342 [astro-ph.CO]}
  \BibitemShut {NoStop}%
\bibitem [{\citenamefont {Aggarwal}\ \emph {et~al.}(2025)\citenamefont
  {Aggarwal} \emph {et~al.}}]{DAMIC-M:2025luv}%
  \BibitemOpen
  \bibfield  {author} {\bibinfo {author} {\bibfnamefont {K.}~\bibnamefont
  {Aggarwal}} \emph {et~al.} (\bibinfo {collaboration} {DAMIC-M}),\ }\bibfield
  {title} {\bibinfo {title} {{Probing Benchmark Models of Hidden-Sector Dark
  Matter with DAMIC-M}},\ }\href {https://doi.org/10.1103/2tcc-bqck} {\bibfield
   {journal} {\bibinfo  {journal} {Phys. Rev. Lett.}\ }\textbf {\bibinfo
  {volume} {135}},\ \bibinfo {pages} {071002} (\bibinfo {year} {2025})},\
  \Eprint {https://arxiv.org/abs/2503.14617} {arXiv:2503.14617 [hep-ex]}
  \BibitemShut {NoStop}%
\bibitem [{\citenamefont {Aguilar-Arevalo}\ \emph {et~al.}(2022)\citenamefont
  {Aguilar-Arevalo} \emph {et~al.}}]{Oscura:2022vmi}%
  \BibitemOpen
  \bibfield  {author} {\bibinfo {author} {\bibfnamefont {A.}~\bibnamefont
  {Aguilar-Arevalo}} \emph {et~al.} (\bibinfo {collaboration} {Oscura}),\
  }\bibfield  {title} {\bibinfo {title} {{The Oscura Experiment}},\ }\href@noop
  {} {\  (\bibinfo {year} {2022})},\ \Eprint {https://arxiv.org/abs/2202.10518}
  {arXiv:2202.10518 [astro-ph.IM]} \BibitemShut {NoStop}%
\bibitem [{\citenamefont {Tiffenberg}\ \emph {et~al.}(2017)\citenamefont
  {Tiffenberg}, \citenamefont {Sofo-Haro}, \citenamefont {Drlica-Wagner},
  \citenamefont {Essig}, \citenamefont {Guardincerri}, \citenamefont {Holland},
  \citenamefont {Volansky},\ and\ \citenamefont {Yu}}]{Tiffenberg:2017aac}%
  \BibitemOpen
  \bibfield  {author} {\bibinfo {author} {\bibfnamefont {J.}~\bibnamefont
  {Tiffenberg}}, \bibinfo {author} {\bibfnamefont {M.}~\bibnamefont
  {Sofo-Haro}}, \bibinfo {author} {\bibfnamefont {A.}~\bibnamefont
  {Drlica-Wagner}}, \bibinfo {author} {\bibfnamefont {R.}~\bibnamefont
  {Essig}}, \bibinfo {author} {\bibfnamefont {Y.}~\bibnamefont {Guardincerri}},
  \bibinfo {author} {\bibfnamefont {S.}~\bibnamefont {Holland}}, \bibinfo
  {author} {\bibfnamefont {T.}~\bibnamefont {Volansky}},\ and\ \bibinfo
  {author} {\bibfnamefont {T.-T.}\ \bibnamefont {Yu}} (\bibinfo {collaboration}
  {SENSEI}),\ }\bibfield  {title} {\bibinfo {title} {{Single-electron and
  single-photon sensitivity with a silicon Skipper CCD}},\ }\href
  {https://doi.org/10.1103/PhysRevLett.119.131802} {\bibfield  {journal}
  {\bibinfo  {journal} {Phys. Rev. Lett.}\ }\textbf {\bibinfo {volume} {119}},\
  \bibinfo {pages} {131802} (\bibinfo {year} {2017})},\ \Eprint
  {https://arxiv.org/abs/1706.00028} {arXiv:1706.00028 [physics.ins-det]}
  \BibitemShut {NoStop}%
\bibitem [{\citenamefont {Bloch}\ \emph {et~al.}(2025)\citenamefont {Bloch}
  \emph {et~al.}}]{SENSEI:2024yyt}%
  \BibitemOpen
  \bibfield  {author} {\bibinfo {author} {\bibfnamefont {I.~M.}\ \bibnamefont
  {Bloch}} \emph {et~al.} (\bibinfo {collaboration} {SENSEI}),\ }\bibfield
  {title} {\bibinfo {title} {{SENSEI at SNOLAB: Single-Electron Event Rate and
  Implications for Dark Matter}},\ }\href
  {https://doi.org/10.1103/PhysRevLett.134.161002} {\bibfield  {journal}
  {\bibinfo  {journal} {Phys. Rev. Lett.}\ }\textbf {\bibinfo {volume} {134}},\
  \bibinfo {pages} {161002} (\bibinfo {year} {2025})},\ \Eprint
  {https://arxiv.org/abs/2410.18716} {arXiv:2410.18716 [astro-ph.CO]}
  \BibitemShut {NoStop}%
\bibitem [{\citenamefont {Salucci}\ \emph {et~al.}(2010)\citenamefont
  {Salucci}, \citenamefont {Nesti}, \citenamefont {Gentile},\ and\
  \citenamefont {Martins}}]{Salucci:2010qr}%
  \BibitemOpen
  \bibfield  {author} {\bibinfo {author} {\bibfnamefont {P.}~\bibnamefont
  {Salucci}}, \bibinfo {author} {\bibfnamefont {F.}~\bibnamefont {Nesti}},
  \bibinfo {author} {\bibfnamefont {G.}~\bibnamefont {Gentile}},\ and\ \bibinfo
  {author} {\bibfnamefont {C.~F.}\ \bibnamefont {Martins}},\ }\bibfield
  {title} {\bibinfo {title} {{The dark matter density at the Sun's location}},\
  }\href {https://doi.org/10.1051/0004-6361/201014385} {\bibfield  {journal}
  {\bibinfo  {journal} {Astron. Astrophys.}\ }\textbf {\bibinfo {volume}
  {523}},\ \bibinfo {pages} {A83} (\bibinfo {year} {2010})},\ \Eprint
  {https://arxiv.org/abs/1003.3101} {arXiv:1003.3101 [astro-ph.GA]}
  \BibitemShut {NoStop}%
\bibitem [{\citenamefont {Pospelov}\ \emph
  {et~al.}(2008{\natexlab{b}})\citenamefont {Pospelov}, \citenamefont {Ritz},\
  and\ \citenamefont {Voloshin}}]{Pospelov:2008jk}%
  \BibitemOpen
  \bibfield  {author} {\bibinfo {author} {\bibfnamefont {M.}~\bibnamefont
  {Pospelov}}, \bibinfo {author} {\bibfnamefont {A.}~\bibnamefont {Ritz}},\
  and\ \bibinfo {author} {\bibfnamefont {M.~B.}\ \bibnamefont {Voloshin}},\
  }\bibfield  {title} {\bibinfo {title} {{Bosonic super-WIMPs as keV-scale dark
  matter}},\ }\href {https://doi.org/10.1103/PhysRevD.78.115012} {\bibfield
  {journal} {\bibinfo  {journal} {Phys. Rev. D}\ }\textbf {\bibinfo {volume}
  {78}},\ \bibinfo {pages} {115012} (\bibinfo {year} {2008}{\natexlab{b}})},\
  \Eprint {https://arxiv.org/abs/0807.3279} {arXiv:0807.3279 [hep-ph]}
  \BibitemShut {NoStop}%
\end{thebibliography}%

\end{document}